\documentclass[12pt]{article}
\usepackage[dvips]{color}
\usepackage{epsfig}
\usepackage{amsmath}
\usepackage{graphicx}

\begin{document}
\title{\bf Drag force with different charges in STU background and $AdS$/CFT}
\author{{J. Sadeghi $^{a}$\thanks{Email: pouriya@ipm.ir}\hspace{1mm}
M. R. Setare $^{b}$ \thanks{Email: rezakord@ipm.ir} and B.
Pourhassan$^{a}$}\thanks{Email: b.pourhassan@umz.ac.ir}\\
{ $^{a}$ Sciences Faculty, Department of Physics, Mazandaran
University,}\\{  P .O .Box 47415-416, Babolsar, Iran} \\ { $^b$
Department of Science, Payame Noor University, Bijar, Iran}}
\maketitle
\begin{abstract}
In this paper we consider the drag force problem in the STU model.
Already this problem and related topics were studied in ${\mathcal
N}=4$ SYM theory. Here we study the same problem in ${\mathcal N}=2$
supergravity theory with an R-charge black hole (the STU model).
This paper extends our previous work [1,2] by studying more general
charge configurations and by taking into account the effect of a
constant electromagnetic field and higher derivative corrections. We
consider a single quark and a $q \bar q$ pair in a three charge
non-extremal black hole background. First by using the AdS/CFT
correspondence, we obtain the drag force, the diffusion coefficient,
the total energy and momentum, quasinormal modes, and the effect of
higher derivative corrections for the single quark. Next we
calculate the drag force for a $q \bar q$ pair which rotates around
its center of mass. For the first time we study the motion of a
rotating $q \bar q$ pair in this background. We show that our
results in the near-extremal limit and finit chemical potential
agree with the energy loss of the moving quark which was calculated
in the
${\mathcal N}=4$ SYM plasma.\\\\
\noindent {\bf Keywords:} $AdS$/CFT correspondence; STU model;
String theory; Drag Force\\\\
\end{abstract}
\section{Introduction}
The formulation of QCD as a string theory plays important role in
the theoretical physics. In the other hand the  $AdS$/CFT
correspondence [3-9] prepares useful methods for the studying QCD in
the strong coupling. Recently, the problem of QCD from $AdS$/CFT
correspondence is discussed by several papers. This problem is the
motion of a heavy charged particle through a thermal medium. Already
this problem considered in the ${\mathcal{N}}$=4 super Yang-Mills
(SYM) thermal plasma and well studied in many papers such as
[10-16], the basis of these papers was a toy model. In the CFT side
there is a moving quark through the ${\mathcal{N}}$=4 SYM thermal
plasma with the momentum $P$, mass $m$ and constant velocity $v$,
which is influenced by an external force $F$. So, one can write the
equation of motion as $\dot{P}=F-\mu P$, where in the
non-relativistic motion $P=mv$ and in the relativistic motion
$P=\frac{mv}{\sqrt{1-v^{2}}}$, also $\mu$ is called friction
coefficient. In that case, one can find the friction coefficient
$\mu$ as a velocity independent [14] or velocity dependent [11]. In
order to obtain drag force, one can consider two special cases. The
first case is the constant momentum ($\dot{P}=0$), so for the
non-relativistic motion one can obtain $F=(\mu m)v$. In this case
the drag force coefficient $(\mu m)$ will be obtained. In the second
case, external force is zero, so one can find $P(t)=P(0)exp(-\mu
t)$. In other words, by measuring the ratio $\frac{\dot{P}}{P}$ or
$\frac{\dot{v}}{v}$ one can determine friction coefficient $\mu$
without any dependence to mass $m$. These methods lead us to obtain
the drag force for a moving quark in the thermal plasma. In the
other hand in $AdS$ side there is type II string theory in
$AdS_{5}\times S^{5}$ space. So, instead of the quark in the gauge
theory, there is an open string in $AdS$ space which
stretched from D-brane to black hole horizon.\\
By using $AdS$/CFT correspondence the rate of energy loss of a heavy
moving quark through  the ${\mathcal{N}}$=4 SYM thermal plasma at
large 't Hooft coupling are determined  by [10-16]. Also there are
many calculations to obtain the jet-quenching parameter [7-24]. The
jet-quenching parameter of quarks is one of the interesting
properties of the strongly-coupled plasma. The jet-quenching
parameter controls the description of relativistic heavy quarks.
Above considerations may be generalized to the case of a
quark-antiquark pair which moves with the constant velocity $v$
through a strongly-coupled thermal ${\mathcal{N}}$=4 SYM plasma
[25-30]. Ref. [27] found the $q\bar{q}$ pair feels no drag force.
Actually the $q\bar{q}$ pair may have more degrees of freedom such
as the rotational motion and oscillation along the connection axis.
In the Refs. [28, 30] the description of $q\bar{q}$ system instead
single quark well explained. Also the problem of spining open string
(meson) in description of non-critical string/gauge duality
considered [31]. In that paper the relation between the energy and
angular momentum of spining open string for the Regge trajectory of
mesons in a QCD-like theory is studied [32, 33]. In this paper we
would like to add rotational motion to the $q\bar{q}$ pair and
calculate drag force. In that case
our method is different with Refs. [31, 34].\\
Then, it is possible to consider higher derivative corrections to
the black hole solution. In that case the effect of
curvature-squared corrections on the drag force of moving heavy
quark in the ${\mathcal{N}}$=4 SYM plasma is considered [35] and
found that curvature-squared corrections may increases or
decreases the value of drag force on the quark.\\
The main subject of this work is to study moving quark and
quark-antiquark pair through ${\mathcal{N}}$=2 supergravity plasma
in the general form, i.e. five dimensional black hole with three
$U(1)$ charges. We will begin with free string in the AdS space and
obtain drag force. This force, in the absence of any external field,
decreases the velocity of string, so, after long time, the motion of
string will be small fluctuations around static string. But in the
presence of appropriate external field (for example, constant
electromagnetic field) string can remain in constant speed. In this
paper, we consider a general STU model with non-zero charges which
is the ${\mathcal{N}}$=2 supergravity theory and has non-extremal
black hole. The solutions of ${\mathcal{N}}$=2 supergravity may be
solutions of supergravity theory with more supersymmetry such as
${\mathcal{N}}$=4 and ${\mathcal{N}}$=8. Indeed, ${\mathcal{N}}$=8
supergravity correspond to the dual ${\mathcal{N}}$=4 SYM. Also
${\mathcal{N}}$=4 supergravity correspond to ${\mathcal{N}}$=2 SCFT
[36, 37], and ${\mathcal{N}}$=2 supergravity correspond to
${\mathcal{N}}$=1 SCFT. Already the duality between gravity and
${\mathcal{N}}$=2 gauged theory investigated [38, 39]. The
${\mathcal{N}}$=2 supergravity theory in five dimensions can be
obtained by compactifying eleven dimensional supergravity in a
three-fold Calabi-Yau [40]. In order to consider effect of
$R^{2}$-term in the curvature tensor and higher derivative
corrections to the ${\mathcal{N}}$=2 supergravity theory we used
from results of Refs. [41, 42], and obtain modified drag force. It
is important to note that, above analysis might be invalid in the
lower dimension with higher derivative effective action without
consideration of proper Kaluza-Klein reductional dimension
[43].\\
The aim of this paper is to study the drag force problem in STU
model with three-charge non-extremal black hole (with different
non-zero charges). Already we studied the drag force on moving quark
at the ${\mathcal{N}}$=2 supergravity and STU model in two papers
[1, 2]. In the first paper [1], we considered a moving quark in the
thermal plasma at the ${\mathcal{N}}$=2 supergravity theory and
obtained the drag force on the quark and quasi-normal modes of the
string. Also, we calculated the effect of higher derivative terms
and existence of external NSNS B-field on the drag force. In that
paper we found the drag force problem in the ${\mathcal{N}}$=2
supergravity theory at the $\eta\rightarrow0$ limit correspond to
drag force problem in the ${\mathcal{N}}$=4 SYM plasma, where $\eta$
is the non-extremality parameter. Then in the second paper [2], we
considered a moving quark in $D=5$, ${\mathcal{N}}$=2 supergravity
plasma and used three-charge non-extremal black hole solution (STU
solution) with only one
non-zero charge and calculated drag force on the quark and quasi-normal modes of the string.\\
Now, we generalize our previous papers to the case of three non-zero
charges ($q_{1}\neq q_{2}\neq q_{3}\neq 0$) and calculate the drag
force on the quark-antiquark pair with linear motion with constant
speed $v$ and circular motion with angular velocity $\omega$. In
that case if we set $q_{1}=q$ and $q_{2}=q_{3}=0$ our problem
reduces to the Ref. [2] and if we set
$q_{1}=q_{2}=q_{3}=q=\eta\sinh^{2}\beta$
then our problem reduces to the Ref. [1].\\
The remaining part of the paper is organized as following. In the
section 2 we briefly review the STU model and calculating method of
the drag force in three-charge non-extremal black hole background.
Then in the section 3 we calculate the drag force for single quark
moving through ${\mathcal{N}}$=2 thermal plasma. The quasi-normal
modes of the curved string and corresponding energy and momentum are
obtained in the section 4. In the section 5 the effect of constant
electromagnetic field, as the external field, on the drag force is
discussed. In section 6, first we consider the quark-antiquark pair
moving with constant speed and then add a rotating motion to obtain
drag force for these configurations. Finally, in the section 7 we
find effect of higher derivative correction to the drag force for
the single quark solution, and in the section 8 we summarize our
results and giving several suggestions for the future works.
\section{STU Model Solution and Drag Force}
The STU model is just some ${\mathcal{N}}$=2 supergravity, which has
generally 8-charge (4 electric and 4 magnetic) non-extremal black
hole. But there are many situations with less than charges such as
four-charge and three-charge black hole. In that case there is great
difference between the three-charge and four-charge black hole. For
example if there are only 3 charges the entropy vanishes (except in
the non-BPS case). So, one really needs four charges to get a
regular black hole. In 5 dimensions the situation is different and
actually much simpler, there is no distinction between BPS and
non-BPS branch. So, in 5 dimensions the three-charge configurations
are the most interesting ones [44]. Therefore, we begin with the
three-charge non-extremal black hole solution in ${\mathcal{N}}$=2
gauged supergravity which is called STU model and described by the
following solution [39, 45],
\begin{equation}\label{s1}
ds^{2}=-\frac{f_{k}}{{\mathcal{H}}^{\frac{2}{3}}}dt^{2}
+{\mathcal{H}}^{\frac{1}{3}}(\frac{dr^{2}}{f_{k}}+\frac{r^{2}}{L^{2}}dx^{2}),
\end{equation}
where,
\begin{eqnarray}\label{s2}
f_{k}&=&k-\frac{m_{k}}{r^{2}}+\frac{r^{2}}{L^{2}}{\mathcal{H}},\nonumber\\
{\mathcal{H}}&=&\prod_{i=1}^{3} H_{i},\nonumber\\
H_{i}&=&1+\frac{q_{i}}{r^{2}}, \hspace{10mm} i=1, 2, 3,
\end{eqnarray}
where $L$ is the constant $AdS$ radius and $r$ is radial coordinate
along the black hole, so the boundary of $AdS$ space is at
$r\rightarrow\infty$ (on the D-brane). And the black hole horizon
specified by $r=r_{h}$ which is obtained from $f_{k}=0$. In STU
model there are three real scalar fields as
$X^{i}=\frac{{\mathcal{H}}^{\frac{1}{3}}}{H_{i}}$ which satisfy $
\prod_{i=1}^{3}X^{i}=1$. For the three R-charges $q_{i}$ in the
equation (2) there is an overall factor such that
$q_{i}=\eta\sinh^{2}\beta_{i}$, where $\eta$ is non-extremality
parameter and $\beta_{i}$ are related to the three independent
electric charges of the black hole. Finally, the factor of $k$
indicates the space curvature, so the metric (1) includes a $S^{3}$
(three dimensional sphere) for $k=1$, a pseudo-sphere for $k=-1$ and
a flat space for $k=0$. The metric (1) written with the assumption
that the motion is only on the transverse axis which is specified by
$x$. We would like to consider a moving charged particle through the
thermal plasma in the above background. In the CFT side there are
complicated calculations to obtain drag force on charged particle.
Therefore we are going to the $AdS$ side. In that case string theory
give us useful mathematical tools to calculate the drag force.
According to the Maldacena dictionary we have stretched string from
D-brane to the black hole horizon in the $AdS$ space. The end point
of string on D-brane represents the charged particle. Also from
string theory it is known that the temperature in the supersymmetric
gauge theory is equal to the existence of a black hole (black brane)
with a flat horizon in the center of the $AdS$ space.\\
In the dual picture of QCD, instead charged particle, we have an
open string which is described by the Nambu-Goto action,
\begin{equation}\label{s3}
S=-\sqrt{-g}=-T_{0}\int{d\tau d\sigma \sqrt{-g}},
\end{equation}
where $T_{0}$ is the string tension and $\tau$ and $\sigma$ are the
string world-sheet coordinates, also $g$ is determinant of the
world-sheet metric $g_{ab}$. We use static gauge, where one fixes
$\tau=t$ and $\sigma=r$. Therefore the string world-sheet is
described by $x(r, t)$, so in order to write lagrangian density one
can find,
\begin{equation}\label{s4}
-g=\frac{1}{{\mathcal{H}}^{\frac{1}{3}}}
\left[1-\frac{{\mathcal{H}}r^{2}}{f_{k}L^{2}}\dot{x}^{2}+\frac{f_{k}r^{2}}{L^{2}}{x^{\prime}}^{2}\right],
\end{equation}
where dot and prime denote $t$ and $r$ derivative respectively. Also
the lagrangian density is given by ${\mathcal{L}}=-T_{0}\sqrt{-g}$.
Then, by using Euler-Lagrange equation one can obtain the string
equation of motion as the following expression,
\begin{equation}\label{s5}
\frac{\partial}{\partial
r}(\frac{f_{k}r^{2}}{{\mathcal{H}}^{\frac{1}{3}}\sqrt{-g}}x^{\prime})=\frac{{\mathcal{H}}^{\frac{2}{3}}r^{2}}{f_{k}}\frac{\partial}{\partial
t}(\frac{\dot{x}}{\sqrt{-g}}),
\end{equation}
where $\sqrt{-g}$ is given by square of the relation (4). In order
to obtain the total energy and momentum, drag force or energy loss
of particle in the thermal plasma, we have to calculate the
canonical momentum densities. In that case one can obtain the
following expressions,
\begin{eqnarray}
\left(\begin{array}{ccc}
\pi_{x}^{0} & \pi_{x}^{1}\\
\pi_{r}^{0}& \pi_{r}^{1}\\
\pi_{t}^{0} & \pi_{t}^{1}\\
\end{array}\right)=-\frac{T_{0}}{{\mathcal{H}}^{\frac{1}{3}}\sqrt{-g}} \left(\begin{array}{ccc}
-\frac{{\mathcal{H}}r^{2}}{f_{k}L^{2}}\dot{x} & \frac{f_{k}r^{2}}{L^{2}}x^{\prime}\\
\frac{{\mathcal{H}}r^{2}}{f_{k}L^{2}}\dot{x}x^{\prime} & 1-\frac{{\mathcal{H}}r^{2}}{f_{k}L^{2}}{\dot{x}}^{2}\\
1+f_{k}\frac{r^{2}}{L^{2}}{x^{\prime}}^{2} & -\frac{f_{k}r^{2}}{L^{2}}\dot{x}x^{\prime}\\
\end{array}\right).
\end{eqnarray}
After that, by using the relation $\dot{P}\propto\pi_{x}^{1}$ we can
obtain the drag force. Here, one can check that above results are
agree with Herzog's works (specially Ref. [10]) at near extremal
limit. In this paper we would like to consider two cases: first, a
moving quark, as a charged particle, through the ${\mathcal{N}}$=2
supergravity thermal plasma which is main subject of sections 3-5,
and second, a moving quark-antiquark
pair in the same background which is considered in the section 6.\\
For the single quark in CFT side, we have an open string in $AdS$
space which stretched from $r=r_{m}$ on D-brane to $r=r_{h}$ at the
horizon. In that case the total energy and momentum of string are
obtained by the following integrals,
\begin{eqnarray}\label{s7}
E&=&-\int_{r_{h}}^{r_{m}}{\pi_{t}^{0}dr},\nonumber\\
P&=&\int_{r_{h}}^{r_{m}}{\pi_{x}^{0}dr}.
\end{eqnarray}
We will use above relations for single quark in the next section.
Also the Hawking temperature of the black hole solution (1) will be
as [45],
\begin{equation}\label{s8}
T_{H}=\frac{r_{h}}{2\pi
L^{2}}\frac{2+\frac{1}{r_{h}^{2}}\sum_{i=1}^{3}{q_{i}}-\frac{1}{r_{h}^{6}}\prod_{i=1}^{3}{q_{i}}}{{\sqrt{\prod_{i=1}^{3}(1+\frac{q_{i}}{r_{h}^{2}})}}}.
\end{equation}
In the following section we will use equation (8) and obtain the
diffusion coefficient for the quark and discuss about drag force of
the single quark.
\section{Single Quark Solution}
As we know a quark in the ${\mathcal{N}}$=2 supergravity thermal
plasma is equal to the stretched string from $r=r_{h}$ on D-brane to
the black hole horizon. Indeed  the D-brane covers an sphere $S^{3}$
in $S^{5}$ space which has minimum radius $r_{m}$, so the string
lives on $AdS_{5}\times S^{5}$ space. Therefore in solutions (1) and
(2) we take $k=1$ and interpret $m_{1}$ as the non-extremality
parameter $\eta$ [2], so from equation (2) we have
$f(r)=1-\frac{\eta}{r^{2}}+\frac{r^{2}}{L^{2}}{\mathcal{H}}$. Hence
from now we set $k=1$ to have $AdS_{5}\times S^{5}$ space [2].\\
Here, there are special cases such as $q_{2}=q_{3}=0$ and $q_{1}=q$
which already discussed [2]. And the case of
$q_{1}=q_{2}=q_{3}=\eta\sinh^{2}\beta$ and
$L^{2}=\frac{1}{\Lambda^{2}}$, where $\Lambda$ is the cosmological
constant, reduce to result of the Ref. [1]. In this paper we take
most general case with $q_{1}\neq q_{2}\neq q_{3}\neq0$. Now, we are
going to discuss about single quark solution and try to obtain drag
force.\\
There is the simplest solution for equation of motion (5), namely
$x=x_{0}$, where $x_{0}$ is a constant and the string stretched
straightforwardly from D-brane at $r=r_{m}$ to the horizon at
$r=r_{h}$. It means that in the dual picture there is a static quark
in the thermal plasma. For such configuration one can obtain
$-g=(H_{1}H_{2}H_{3})^{-\frac{1}{3}}$ and
$\pi_{x}^{0}=\pi_{r}^{0}=\pi_{t}^{1}=\pi_{x}^{1}=0$, so drag force
is zero, as it expected for the static quark. Only non-zero
components of momentum density are
$\pi_{r}^{1}=\pi_{t}^{0}=-T_{0}[\prod_{i=1}^{3}(1+\frac{q_{i}}{r_{h}^{2}})]^{-\frac{1}{6}}$,
so total energy of the string is obtained as,
\begin{equation}\label{s9}
E=T_{0}\left[r+\frac{1}{6r}\sum_{i}{q_{i}}+\frac{1}{36r^{3}}\sum_{i\neq
j}{q_{i}q_{j}}+\frac{1}{30r^{5}}\prod_{i}q_{i}\right]_{r_{h}}^{r_{m}},
\end{equation}
where we assume that the black hole charges $q_{i}$ are small. In
zero temperature one can interpret $E$ as the rest mass of the
quark, which is obtained by the following expression,
\begin{equation}\label{s10}
M_{rest}=T_{0}\left[r_{m}-\frac{16}{15}r_{h}+(\frac{1}{6r_{m}}-\frac{1}{5r_{h}})\sum_{i}{q_{i}}+(\frac{1}{r_{m}^{3}}-\frac{1}{r_{h}^{3}})\frac{\sum_{i\neq
j}{q_{i}q_{j}}}{36r^{3}}+\frac{1}{30r_{m}^{5}}\prod_{i}q_{i}\right].
\end{equation}
In STU model there is non-extremal black hole which is described by
non-extremality parameter, but in the ${\mathcal{N}}$=4 super
Yang-Mills theory there is near-extremal black hole. So, if we take
$\eta\rightarrow0$ limit, we have near-extremal black hole, then the
total energy of string obtained as $E=T_{0}(r_{m}-r_{h})$ [10], so,
the zero temperature  is equal $r_{h}=0$ and the physical mass of
quark (rest mass) become $M_{rest}=T_{0}r_{m}$. In the other hand if
we set $q_{1}=q_{2}=q_{3}=q$, the total energy of string is agree
with equation (3.10) of Ref. [1] (under assumption of small
charge with $q=\eta\sinh^{2}\beta$).\\
As we told the total energy of string (9) obtained for static quark.
Now we are going to consider most physical time-dependent solution
of moving quark through plasma which in dual picture is a curved
string described by $x(r,t)=x(r)+vt$, where $v$ is the constant
velocity of the single quark [1, 2, 10]. In that case by using
equation of motion (5) one can find,
\begin{equation}\label{s11}
\frac{f(r)r^{2}}{L^{2}v{\mathcal{H}}^{\frac{1}{3}}\sqrt{-g}}x^{\prime}=C,
\end{equation}
where $C$ is an integration constant and $\sqrt{-g}$ is obtained in
the following equation,
\begin{equation}\label{s12}
-g=\frac{1}{{\mathcal{H}}^{\frac{1}{3}}}
\left[1-\frac{{\mathcal{H}}r^{2}}{f(r)L^{2}}v^{2}+\frac{f(r)r^{2}}{L^{2}}{x^{\prime}}^{2}\right].
\end{equation}
Therefore it is easy to check that
$\pi_{t}^{1}=-\pi_{x}^{1}v=T_{0}Cv^{2}$. These constant components
of momentum density is obtained already for ${\mathcal{N}}$=4 SYM
theory [1, 10, 11]. Now, by combining equations (11) and (12) one
can obtain,
\begin{equation}\label{s13}
-g=\frac{r^{2}}{{\mathcal{H}}^{\frac{1}{3}}}
\frac{f(r)-{\mathcal{H}}r^{2}v^{2}}{f(r)r^{2}-C^{2}v^{2}{\mathcal{H}}^{\frac{1}{3}}}.
\end{equation}
Reality condition for $\sqrt{-g}$ and therefor $x^{\prime}$ tell us
that we should choose,
\begin{equation}\label{s14}
C=\left[\prod_{i=1}^{3}(1+\frac{q_{i}}{r_{c}^{2}})\right]^{\frac{1}{3}}r_{c}^{2},
\end{equation}
to have real energy and momentum. In the equation (14) critical
radius $r_{c}$ is the root of the following equation,
\begin{equation}\label{s15}
r^{6}+\mathcal{A}r^{4}+\mathcal{B}r^{2}+\prod_{i=1}^{3}q_{i}=0,
\end{equation}
where $\mathcal{A}=\sum_{i=1}^{3}{q_{i}}+\frac{L^{2}}{1-L^{2}v^{2}}$
and $\mathcal{B}=\frac{1}{2}\sum_{i\neq j}q_{i}q_{j}-\frac{\eta
L^{2}}{1-L^{2}v^{2}}$. Thus, one can find the drag force on single
moving quark as,
\begin{equation}\label{s16}
\dot{P}=-T_{0}\pi_{x}^{1}=-T_{0}\left[\prod_{i=1}^{3}(1+\frac{q_{i}}{r_{c}^{2}})\right]^{\frac{1}{3}}vr_{c}^{2}.
\end{equation}
Indeed the equation (16) is the momentum current into the horizon.
Here, as already discussed [11], we have field theory interpretation
of our system. One can image single quark moving in a constant
external field with strength
$\varepsilon=T_{0}\left[\prod_{i=1}^{3}(1+\frac{q_{i}}{r_{c}^{2}})\right]^{\frac{1}{3}}vr_{c}^{2}$.
This external field keeps the curved string moving at the constant
speed $v$. We know that electromagnetic field lives on a D-brane on
which this dragging string ends. The $\varepsilon$ changes the
boundary conditions for the string. Usually, the string should
satisfy Dirichlet boundary conditions orthogonal to the D-brane and
Neumann boundary conditions parallel ($\pi_{x} = 0$). In the
presence of $\varepsilon$, the Neumann boundary conditions can be
altered. The fact that $\pi_{x} \neq 0$ for the string solution
indicates the presence of an electromagnetic field on the D-brane.
We will study this statement in detail in the section 5. Also by
using relations $\dot{P}=-\mu m v$, $D=\frac{T_{H}}{\mu m}$, (8) and
(16) one can obtain diffusion
coefficient ($D$) of the quark.\\
Now, final expression for $x^{\prime}(r)$ is,
\begin{equation}\label{s17}
x^{\prime}(r)^{2}=\alpha\frac{{\mathcal{H}}(r)^{\frac{2}{3}}}{f(r)^{2}r^{4}},
\end{equation}
where $\alpha=v^{2}L^{2}r_{h}^{4}{\mathcal{H}}(r_{h})^{\frac{1}{3}}$
is a constant. Then by using equations (6) and (17) we arrive to the
following relations,
\begin{eqnarray}\label{s18}
\pi_{t}^{0}&=&-\gamma\frac{1+\frac{\alpha{\mathcal{H}}(r)^{\frac{2}{3}}}{f(r)r^{2}L^{2}}}{{\mathcal{H}}(r)^{\frac{1}{3}}},\nonumber\\
\pi_{x}^{0}&=&\gamma\frac{v}{L^{2}}\frac{{\mathcal{H}}(r)^{\frac{2}{3}}r^{2}}{f(r)},
\end{eqnarray}
where
$\gamma=T_{0}(\frac{r_{c}}{r_{h}})^{2}\frac{{\mathcal{H}}(r_{c})^{\frac{1}{3}}}{{\mathcal{H}}(r_{h})^{\frac{1}{6}}}$
is another constant. Then by using equations (7) and (18) one can
obtain total energy and momentum of the string.\\
It is interesting to consider the relativistic motion. In that case
$P=\frac{mv}{\sqrt{1-v^{2}}}$ and therefore friction coefficient
$\mu$ may be expressed as the following,
\begin{equation}\label{s19}
\mu
m=T_{0}\left[\prod_{i=1}^{3}(1+\frac{q_{i}}{r_{c}^{2}})\right]^{\frac{1}{3}}\sqrt{1-v^{2}}r_{c}^{2}.
\end{equation}
Now in the case $q_{1}=q_{2}=q$ and $q_{3}=0$ we have,
\begin{equation}\label{s20}
\mu
m=\frac{T_{0}}{2}\sqrt{1-v^{2}}(1+\frac{q}{r_{c}^{2}})^{\frac{2}{3}}
\left[2q+\frac{1}{1-v^{2}}\pm\sqrt{(2q+\frac{1}{1-v^{2}})^{2}-4(\frac{q^{2}}{2}-\frac{\eta}{1-v^{2}})}\right],
\end{equation}
for simplicity we set $L^{2}=1$. Relation (20) for small $q$ reduces
to,
\begin{equation}\label{s21}
\mu
m=T_{0}q\sqrt{1-v^{2}}\left[1+\frac{1}{3(1-v^{2})r_{c}^{2}}\right]
\left[1\pm1+\frac{2\eta(1-v^{2})}{1+4q(1-v^{2})}\right]+{\mathcal{O}}(q^{2}),
\end{equation}
Also in the near-extremal limit $\eta\rightarrow0$ we have $\mu
m=\frac{T_{0}}{\sqrt{1-v^{2}}}$ in the relativistic motion and $\mu
m=\frac{T_{0}}{1-v^{2}}$ for the non-relativistic motion. In that
case we used positive sign in relation (20) because negative sign
yield to zero friction coefficient in $\eta\rightarrow0$ limit.
Therefore we should take minus sign in relations (20) and (21).
These relations show that increasing of $q$ increase the value of
$\mu m$, and therefore value of drag force. The drag force
coefficient $\mu m$ in the relation (21) has a maximum at the
following value of $q$,
\begin{equation}\label{s22}
q\simeq\frac{1-\eta(1-v^{2})}{4(1-v^{2})\left[2+\eta(1-v^{2})(1-\eta(1-v^{2}))\right]},
\end{equation}
which reduces to $q\simeq\frac{1}{8(1-v^{2})}$ at the near-extremal
limit. In this limit the maximum of the drag force coefficient is,
\begin{equation}\label{s23}
(\mu
m)_{max}=\frac{T_{0}}{4\sqrt{1-v^{2}}}\left[1+\frac{1}{3(1-v^{2})r_{c}^{2}}\right].
\end{equation}
Here, it is interesting to compare our above results to the
calculations of Ref. [11] (specially equation (20) in this paper
with equation (5.9) of Ref. [11]), where SYM in Minkowski space
($k=0$) is considered. It is expected that for YM on a sphere
Lorentz invariance to be broken and one can see that our solutions
should be identical to Ref. [11] at non-relativistic case
($v^{2}\rightarrow0$).\\ In the next section we are going to discuss
about quasi-normal modes. In the section 3 we considered a moving
string through thermal plasma without any external field, and
obtained non-zero drag force. It means that the motion of string
should be reduces to small fluctuations after the long time, which
known as quasi-normal modes of string.
\section{Quasi-Normal Modes of Curved String}
In this section, we consider small perturbations of a curved string
which stretched from $r=r_{m}$ to $r=r_{h}$ in STU background with
three non-zero charges. This consideration allows us to obtain the
friction coefficient $\mu$ in the non-relativistic regime of the
quark. In that case we consider a quark moving in the
${\mathcal{N}}$=2 supergravity thermal plasma without any external
field. Indeed, we want to study the behavior of the curved string at
the $t\rightarrow\infty$ and low velocity limits. We use the same
method with Ref. [11], where small perturbations of a straight
string in ${\mathcal{N}}$=4 SYM thermal plasma considered. For more
detail of such configuration see Refs. [1, 2, 45, 46, 47, 48, 49,
50]. The small fluctuations around the curved string means that
${\dot{x}}^{2}$ and ${x^{\prime}}^{2}$ are small, so one can neglect
them in the expression (4). Hence the equation of motion (5) reduces
to the following equation,
\begin{equation}\label{s24}
\frac{\partial}{\partial
r}\left(\frac{f(r)r^{2}}{{\mathcal{H}}^{\frac{1}{6}}}x^{\prime}\right)
-\frac{{\mathcal{H}}^{\frac{5}{6}}r^{2}}{f(r)}\ddot{x}=0.
\end{equation}
Then under assumption of time-dependent solution of the form $x(r,
t)=x(r) e^{-\mu t}$, equation of motion (24) reduces to the simple
eigenvalue equation $Ox(r)=\mu^{2}x(r)$, where we introduce operator
$O$ as the following,
\begin{equation}\label{s25}
O=\frac{f(r)}{{\mathcal{H}}^{\frac{5}{6}}r^{2}}\frac{d}{dr}\frac{f(r)r^{2}}{{\mathcal{H}}^{\frac{1}{6}}}\frac{d}{dr}.
\end{equation}
In order to obtain friction coefficient, we assume that $\mu$ is
small and one can use expansion $x(r)=x_{0}(r)+\mu
x_{1}(r)+\mu^{2}x_{2}(r)+\cdots$. Also by applying Neumann boundary
condition we find $x^{\prime}(r_{m})=\mu
x_{1}^{\prime}(r_{m})+\mu^{2}x_{2}^{\prime}(r_{m})=0$ and $x_{0}$
specified as a constant. Therefore under assumption of small $\mu$
one can find,
\begin{equation}\label{s26}
\mu=-\frac{A}{x_{0}L^{4}}\left[\frac{\left(L^{2}(r^{2}-\eta)
+{\mathcal{H}}(r)r^{4}\right)^{2}}{{\mathcal{H}}(r)^{\frac{1}{3}}}\int{\frac{{\mathcal{H}}(r)^{\frac{5}{6}}}{L^{2}
(r^{2}-\eta)+{\mathcal{H}}(r)r^{4}}dr}\right]_{r=r_{m}},
\end{equation}
where $A$ is an integration constant. Then, by using the relation
(26) we can determine the value of the drag force. One can check
that the drag force will obtained
as a constant  and at the near-extremal limit it is proportional to the $\tan^{-1}(\frac{r_{m}}{L})$.\\
Now we can use these results to obtain the total energy and momentum
of curved string. In that case we use equation of motion (24) and
relations $\dot{x}=-\mu x$,
$P=\int_{r_{min}}^{r_{m}}{\pi_{x}^{0}dr}$ and also Neumann boundary
condition ($x^{\prime}(r_{m})=0$), so the total momentum of string
will be as the following relation,
\begin{equation}\label{s27}
P=\frac{T_{0}}{\mu
L^{2}}\left[\frac{r_{min}^{6}+(L^{2}+\sum_{i}{q_{i}})r_{min}^{4}+(\frac{1}{2}\prod_{i\neq
j}q_{i}q_{j}-\eta
L^{2})r_{min}^{2}+\sum_{i}q_{i}}{r_{min}^{12}+r_{min}^{6}\sum_{i}q_{i}+\frac{1}{2}r_{min}^{3}\prod_{i\neq
j}q_{i}q_{j}+r_{min}^{2}\sum_{i}q_{i}}\right]x^{\prime}(r_{min}),
\end{equation}
where we insert $r_{min}>r_{h}$ as lower limit of integral to avoid
divergency.\\
In order to obtain the total energy we keep second order of
velocities and expand $\sqrt{-g}$ to find,
\begin{eqnarray}\label{s28}
E&=&-\frac{T_{0}}{2 L^{2}}\left[\frac{r_{min}^{2}L^{2}-\eta L^{2}
+r_{min}^{4}\prod_{i}(1+\frac{q_{i}}{r_{min}^{2}})}{\prod_{i}(1+\frac{q_{i}}{r_{min}^{2}})^{\frac{1}{6}}}x(r_{min})x_{1}^{\prime}(r_{min})\right]\nonumber\\
&-&T_{0}\int_{r_{min}}^{r_{m}}dr\prod_{i}(1+\frac{q_{i}}{r^{2}})^{\frac{1}{6}},
\end{eqnarray}
where for small values of charge reduces to the following
expression,
\begin{eqnarray}\label{s29}
E&\simeq &-\frac{T_{0}}{2
L^{2}}\left[r_{min}^{4}+r_{min}^{2}(L^{2}+\frac{5}{6}\sum_{i}q_{i})+\frac{1}{2}\prod_{i\neq
j}q_{i}q_{j}-\frac{L^{2}}{6}\sum_{i}q_{i}-\eta L^{2}\right]x(r_{min})x^{\prime}(r_{min})\nonumber\\
&-&T_{0}\left(r_{m}-r_{min}+\frac{1}{6}(\frac{1}{r_{m}}-\frac{1}{r_{min}})\sum_{i}q_{i}\right),
\end{eqnarray}
In order to check validity of equations (27) and (29) we take the
near-extremal limit and find that,
\begin{eqnarray}\label{s30}
P&=&\frac{T_{0}}{\mu
L^{2}}r_{min}^{2}(r_{min}^{2}+L^{2})x^{\prime}(r_{min}),\nonumber\\
E&=&T_{0}\left[r_{m}-r_{min}-\frac{1}{2L^{2}}(r_{min}^{2}+L^{2})r_{min}^{2}x(r_{min})x^{\prime}(r_{min})\right],
\end{eqnarray}
which is agree with [1, 10] under assumption
$L^{2}=\frac{1}{\Lambda^{2}}$. Therefore one can write
$E=T_{0}(r_{m}-r_{min})-\frac{\mu P}{2}x(r_{min})$. So, by using
relation $P=m\dot{x}=-m \mu x$, where $m$ is the kinetic mass of the
quark, we have a simple relation between the energy and momentum as
$E=T_{0}(r_{m}-r_{min})-\frac{P^{2}}{2m}$. The first term of right
hand side interpreted as $M_{rest}$ of the quark.
\section{Effect of Constant Electromagnetic Field}
In the previous sections the moving quark through plasma considered
without any external field. In the present section we consider the
single quark which moves with constant speed $v$ through thermal
plasma in STU background, and introduce a constant electromagnetic
field as external field. We assume that the constant electromagnetic
field is along $x^{1}$ and $x^{2}$ directions. Therefore we add a
constant $B$-field in the form $B=B_{01}dt\wedge
dx_{1}+B_{12}dx_{1}\wedge dx_{2}$ to the line element (1), where
$B_{01}$ is the constant electric field and $B_{12}$ is the constant
magnetic field. Also $B_{01}$ and $B_{12}$ are antisymmetric fields
and other components of the $B$-field are zero.
We must note that the same work was done originally for $\mathcal{N}$=4 SYM theory [15].\\
Because of introducing $B_{01}$ and $B_{12}$, the curved string dual
to quark may be described by the $x_{1}(r, t)=x_{1}(r)+v_{1}t$,
$x_{2}(r, t)=x_{2}(r)+v_{2}t$ and $x_{3}(r, t)=0$. Therefore the
square root of lagrangian density (4) takes the following form,
\begin{equation}\label{s31}
-g=\frac{1}{{\mathcal{H}}^{\frac{1}{3}}}\left[1-\frac{{\mathcal{H}}r^{2}}
{f(r)L^{2}}\vec{v}^{2}+\frac{f(r)r^{2}}{L^{2}}{x^{\prime}}^{2}-(B_{01}x_{1}^{\prime}
+B_{12}(\vec{v}\times\vec{x}^{\prime}))^{2}\right],
\end{equation}
where $\vec{v}=(v_{1}, v_{2})$ is the vector of velocity and
$\vec{x}^{\prime}=(x_{1}^{\prime}, x_{2}^{\prime})$ is the projected
directions of string tail. We would like to consider three separated
cases, first, we assume that only electric field is exist and
$B_{12}=0$ and second, we have non-zero magnetic field and there is
$B_{01}=0$, and finally we discuss about the case where $\vec{v}\bot B_{01}$.\\
In order to study effect of constant electric field, one may choose
the moving direction of the quark to be in the $x_{1}$ direction, so
we have $x_{1}(r, t)=x_{1}(r)+vt$ and $x_{2}(r, t)=x_{3}(r, t)=0$.
In that case the $x_{1}$-component of momentum density obtained as,
\begin{equation}\label{s32}
\pi_{x_{1}}^{1}=\sqrt{\frac{\frac{r_{c}^{4}}{L^{2}}\prod_{i}
(1+\frac{q_{i}}{r_{c}^{2}})+r_{c}^{2}-\eta}{(\prod_{i}(1+\frac{q_{i}}{r_{c}^{2}}))^\frac{1}{3}}-B_{01}^{2}},
\end{equation}
where $r_{c}$ is the root of the equation (15). It is easy to check
that if $B_{01}=0$, then the equation (32) reduces to the equation
(16). On the other hand if we choose the value of electric field as
$B_{01}^{2}=\frac{\frac{r_{c}^{4}}{L^{2}}\prod_{i}
(1+\frac{q_{i}}{r_{c}^{2}})+r_{c}^{2}-\eta}{(\prod_{i}(1+\frac{q_{i}}{r_{c}^{2}}))^\frac{1}{3}}$,
then $\pi_{x_{1}}^{1}=0$ and string feels no drag force, therefore
the string can continue its motion at constant velocity. Indeed, it
is what we mentioned after the equation (16).\\
It is clear that at the near-extremal limit the critical radius
reduces to $r_{c}^{2}=\frac{L^{2}}{L^{2}v^{2}-1}$ and we have
$\pi_{x_{1}}^{1}=\sqrt{(\frac{vL^{2}}{L^{2}v^{2}-1})^{2}-B_{01}^{2}}$.
So, if electric field is switched off, then the drag force is
proportional to $\frac{vL^{2}}{L^{2}v^{2}-1}$ which is in agreement
with Refs. [1, 10, 12] under assumption of $L^{2}\Lambda^{2}=1$. In
that case for infinitesimal electric field the drag force modified
by a term in the form of
$\frac{1}{2}T_{0}\frac{v^{2}-1}{v}B_{01}^{2}$. We see that there are
four states where the drag force may be increase or decreases. In
the case of $v>0$, $v^{2}>1$ and $v<0$, $v^{2}<1$ the effect of
existence constant electric field is increasing of the drag force.
But in the case of $v>0$, $v^{2}<1$ and $v<0$, $v^{2}>1$ the
constant electric field decreases the drag force. However in the
relativistic limit $v\rightarrow1$ the constant electric field has
no effect to the drag force.\\
In the second case we want to consider only constant magnetic field
$B_{12}$. In that case we can choose $x_{1}(r, t)=x_{1}(r)+vt$,
$x_{2}(r, t)=x_{2}(r)$ and $x_{3}(r, t)=0$. Under this assumption
one can find,
\begin{eqnarray}\label{s33}
x_{1}^{\prime}(r)&=&\pi_{x_{1}}\left[\frac{\beta(\frac{1}{{\mathcal{H}}^{\frac{1}{3}}}
-\frac{{\mathcal{H}}^{\frac{2}{3}}r^{2}v^{2}}{f(r)})}{\frac{f(r)r^{2}}{{\mathcal{H}}^{\frac{1}{3}}}
\left((\pi_{x_{1}}^{2}-\frac{f(r)r^{2}}{{\mathcal{H}}^{\frac{1}{3}}})(\pi_{x_{2}}^{2}-\beta)
-\pi_{x_{1}}^{2}\pi_{x_{2}}^{2}\right)}\right]^{\frac{1}{2}},\nonumber\\
x_{2}^{\prime}(r)&=&\pi_{x_{2}}\left[\frac{\frac{f(r)r^{2}}{{\mathcal{H}}^{\frac{1}{3}}}(\frac{1}{{\mathcal{H}}^{\frac{1}{3}}}
-\frac{{\mathcal{H}}^{\frac{2}{3}}r^{2}v^{2}}{f(r)})}{\beta
\left((\pi_{x_{1}}^{2}-\frac{f(r)r^{2}}{{\mathcal{H}}^{\frac{1}{3}}})(\pi_{x_{2}}^{2}-\beta)
-\pi_{x_{1}}^{2}\pi_{x_{2}}^{2}\right)}\right]^{\frac{1}{2}},
\end{eqnarray}
where
$\beta=\frac{f(r)r^{2}}{{\mathcal{H}}^{\frac{1}{3}}}-v^{2}B_{12}^{2}$
and we set $\pi_{x_{i}}^{1}\equiv \pi_{x_{i}}$. Now reality
condition implies that $\pi_{x_{2}}=0$ and therefore we have,
\begin{equation}\label{s34}
\pi_{x_{1}}=\left[(1+\frac{q_{1}}{r_{c}^{2}})(1+\frac{q_{2}}{r_{c}^{2}})(1+\frac{q_{3}}{r_{c}^{2}})\right]^{\frac{1}{3}}vr_{c}^{2},
\end{equation}
where $r_{c}$ is the root of equation (15). It tell us that there is
no drag force in $x^{2}$ direction and $B_{12}$ have no effect on
the motion along $x^{1}$ direction since the equation (34) is
coincide with equation (16) which obtained without any external
field. Actually vanishing of $\pi_{x_{2}}$ is consequence of
vanishing of $v_{2}$.\\
With respect to these two cases, (electric and magnetic fields) we
found that the constant magnetic field have no effect on the motion
of string and it is appropriate electric field which keeps the
string at constant speed $v$, so this is agree with results of Refs. [10, 11].\\
Before end of this section we consider the case of $\vec{v}\bot
B_{01}$. It means that one may choose the solutions of equation of
motion as, $x_{1}(r, t)=x_{1}(r)$, $x_{2}(r, t)=x_{2}(r)+vt$ and
$x_{3}(r, t)=0$. A possible drag force may be found as the following
relation,
\begin{equation}\label{s35}
{\dot{P}}_{2}=-T_{0}\sqrt{\frac{\frac{r_{c}^{4}}{L^{2}}\prod_{i}
(1+\frac{q_{i}}{r_{c}^{2}})+r_{c}^{2}-\eta}{(\prod_{i}(1+\frac{q_{i}}{r_{c}^{2}}))^\frac{1}{3}}-v^{2}B_{12}^{2}},
\end{equation}
and ${\dot{P}}_{1}=0$. In this case the constant electric field has
no effect on drag force. We should mention that, because of special
direction of motion which we considered already, this situation is
not happen for our string which studied in the previous sections.
\section{Quark-Antiquark Solutions}
In this section we consider a quark-antiquark pair, which may be
interpreted as a meson, moving with the constant speed $v$ in STU
background. Already the energy of a moving quark-antiquark pair in
${\mathcal{N}}$=4 super Yang-Mills plasma calculated [51]. Now we
would like to repeat same calculation STU background. To represent a
quark-antiquark pair in the dual picture one may consider an open
string in $AdS_{5}$ space which two endpoints of string lie on
D-brane in the ($X, Y$) plan. Two end points of string on the
D-brane represent quark and antiquark which separated from each
other by a constant $l$. We assume that at the $t=0$ string is
straight and two endpoints of string move with the constant velocity
$v$ along the $X$ direction. The dynamics of such configuration
discussed in detail in the Ref. [51]. Similar to the calculations of
the previous sections one can obtain the square root quantity of the
lagrangian density as a following expression,
\begin{equation}\label{s36}
-g=\frac{1}{{\mathcal{H}}^{\frac{1}{3}}}\left[1+\frac{f(r)r^{2}}{L^{2}}({x^{\prime}}^{2}+{y^{\prime}}^{2})
-\frac{{\mathcal{H}}r^{2}}{f(r)L^{2}}(\dot{x}^{2}+\dot{y}^{2})
-\frac{r^{4}}{L^{4}}{\mathcal{H}}(\dot{x}^{2}{y^{\prime}}^{2}+\dot{y}^{2}{x^{\prime}}^{2}-2\dot{x}x^{\prime}\dot{y}y^{\prime})\right],
\end{equation}
and the equations of motion with respect to $x$ and $y$  is given by
the following equations, respectively,
\begin{eqnarray}\label{s37}
\frac{\partial}{\partial
r}\left[\frac{1}{\sqrt{-g}}(\frac{f(r)r^{2}}{{\mathcal{H}}^{\frac{1}{3}}}x^{\prime}
+\frac{r^{4}}{L^{2}}{\mathcal{H}}^{\frac{2}{3}}({\dot{y}}^{2}x^{\prime}-\dot{x}\dot{y}y^{\prime}))\right]\nonumber\\
+r^{2}{\mathcal{H}}^{\frac{2}{3}}\frac{\partial}{\partial
t}\left[\frac{1}{\sqrt{-g}}(\frac{\dot{x}}{f(r)}
+\frac{r^{2}}{L^{2}}({y^{\prime}}^{2}\dot{x}-x^{\prime}\dot{y}y^{\prime}))\right]=0,\nonumber\\
\frac{\partial}{\partial
r}\left[\frac{1}{\sqrt{-g}}(-\frac{f(r)r^{2}}{{\mathcal{H}}^{\frac{1}{3}}}y^{\prime}
+\frac{r^{4}}{L^{2}}{\mathcal{H}}^{\frac{2}{3}}({\dot{x}}^{2}y^{\prime}-\dot{x}\dot{y}x^{\prime}))\right]\nonumber\\
+r^{2}{\mathcal{H}}^{\frac{2}{3}}\frac{\partial}{\partial
t}\left[\frac{1}{\sqrt{-g}}(\frac{\dot{y}L^{2}}{f(r)}
+r^{2}({x^{\prime}}^{2}\dot{y}-x^{\prime}\dot{x}y^{\prime}))\right]=0,
\end{eqnarray}
then momentum densities obtained by following equation,
\begin{eqnarray}\label{s38}
&&\left(\begin{array}{ccc}
\pi_{x}^{0} & \pi_{x}^{1}\\
\pi_{y}^{0} & \pi_{y}^{1}\\
\pi_{r}^{0}& \pi_{r}^{1}\\
\pi_{t}^{0} & \pi_{t}^{1}\\
\end{array}\right)=-T_{0}\frac{r^{2}{\mathcal{H}}^{\frac{1}{3}}}{L^{2}\sqrt{-g}}\times\nonumber\\
&&\left(\begin{array}{ccc}
\frac{r^{2}}{L^{2}}{\mathcal{H}}^{\frac{1}{3}}x^{\prime}\dot{y}y^{\prime}
-(\frac{{\mathcal{H}}^{\frac{1}{3}}}{f(r)}+
\frac{r^{2}}{L^{2}}{\mathcal{H}}^{\frac{1}{3}}{y^{\prime}}^{2})\dot{x}&
\frac{r^{2}}{L^{2}}{\mathcal{H}}^{\frac{1}{3}}y^{\prime}\dot{y}\dot{x}
-(\frac{{\mathcal{H}}^{\frac{1}{3}}r^{2}}{L^{2}}{\dot{y}}^{2}-\frac{f(r)}{{\mathcal{H}}^{\frac{2}{3}}})x^{\prime}\\
\frac{r^{2}}{L^{2}}{\mathcal{H}}^{\frac{1}{3}}y^{\prime}\dot{x}x^{\prime}
-(\frac{{\mathcal{H}}^{\frac{1}{3}}}{f(r)}+
\frac{r^{2}}{L^{2}}{\mathcal{H}}^{\frac{1}{3}}{x^{\prime}}^{2})\dot{y}&
\frac{r^{2}}{L^{2}}{\mathcal{H}}^{\frac{1}{3}}x^{\prime}\dot{y}\dot{x}
-(\frac{{\mathcal{H}}^{\frac{1}{3}}r^{2}}{L^{2}}{\dot{x}}^{2}-\frac{f(r)}{{\mathcal{H}}^{\frac{2}{3}}})y^{\prime}\\
\frac{{\mathcal{H}}^{\frac{1}{3}}}{f(r)}(\dot{x}x^{\prime}+\dot{y}y^{\prime}) &
\frac{L^{2}}{{\mathcal{H}}^{\frac{2}{3}}r^{2}}-\frac{{\mathcal{H}}^{\frac{1}{3}}}{f(r)}({\dot{x}}^{2}+{\dot{y}}^{2})\\
\frac{L^{2}}{{\mathcal{H}}^{\frac{1}{3}}r^{2}}+\frac{f(r)}{{\mathcal{H}}^{\frac{1}{3}}}({x^{\prime}}^{2}+{y^{\prime}}^{2}) &
-\frac{f(r)}{{\mathcal{H}}^{\frac{2}{3}}}(\dot{x}x^{\prime}+\dot{y}y^{\prime})\\
\end{array}\right).
\end{eqnarray}
In this paper we consider two special cases, the first is the moving
quark-antiquark pair with constant speed $v$ and in the second case
we add rotational motion to the pair. The first case may be
described by the $x(r, t) =vt+x(r)$ and $y(r, t)= y(r)$. These
solutions satisfy boundary conditions as $x(\infty, t)=vt$ and
$y(\infty)=\pm\frac{l}{2}$. In this case equation (38) reduces to
the following expression,
\begin{eqnarray}\label{s39}
\left(\begin{array}{ccc}
\pi_{x}^{0} & \pi_{x}^{1}\\
\pi_{y}^{0} & \pi_{y}^{1}\\
\pi_{r}^{0}& \pi_{r}^{1}\\
\pi_{t}^{0} & \pi_{t}^{1}\\
\end{array}\right)=-T_{0}\frac{r^{2}}{L^{2}}\frac{{\mathcal{H}}^{\frac{2}{3}}}{\sqrt{-g}}
\left(\begin{array}{ccc}
-v(\frac{1}{f(r)}+\frac{r^{2}}{L^{2}}{y^{\prime}}^{2})& \frac{f(r)}{{\mathcal{H}}}x^{\prime}\\
\frac{r^{2}}{L^{2}}vy^{\prime}x^{\prime}&-(\frac{r^{2}}{L^{2}}v^{2}-\frac{f(r)}{{\mathcal{H}}})y^{\prime}\\
\frac{v}{f(r)}x^{\prime} &
\frac{L^{2}}{{\mathcal{H}}r^{2}}-\frac{v^{2}}{f(r)}\\
(\frac{L^{2}}{f(r)r^{2}}+{x^{\prime}}^{2}+{y^{\prime}}^{2})\frac{f(r)}{{\mathcal{H}}}
&
-v\frac{f(r)}{{\mathcal{H}}}x^{\prime}\\
\end{array}\right),
\end{eqnarray}
where,
\begin{equation}\label{s40}
-g=\frac{1}{{\mathcal{H}}^{\frac{1}{3}}}\left[1+\frac{f(r)r^{2}}{L^{2}}({x^{\prime}}^{2}+{y^{\prime}}^{2})
-\frac{{\mathcal{H}}r^{2}v^{2}}{f(r)L^{2}}
-\frac{r^{4}}{L^{4}}{\mathcal{H}}v^{2}{y^{\prime}}^{2}\right].
\end{equation}
In order to obtain drag force, we take $\pi_{x}^{1}$ and
$\pi_{y}^{1}$ components and solve them for $x^{\prime}$ and
$y^{\prime}$ respectively and obtain,
\begin{eqnarray}\label{s41}
x^{\prime}(r)&=&\pi_{x}^{1}\frac{L}{r}(1-\frac{{\mathcal{H}}r^{2}v^{2}}{f(r)L^{2}})\left[(\frac{f(r)}{{\mathcal{H}}}-\frac{r^{2}v^{2}}{L^{2}})
(T_{0}^{2}\frac{r^{2}}{L^{2}}f(r){\mathcal{H}}^{\frac{2}{3}}-{\mathcal{H}}{\pi_{x}^{1}}^{2})-f(r){\pi_{y}^{1}}^{2}
\right]^{-\frac{1}{2}},\nonumber\\
y^{\prime}(r)&=&\pi_{y}^{1}\frac{L}{r}\left[(\frac{f(r)}{{\mathcal{H}}}-\frac{r^{2}v^{2}}{L^{2}})
(T_{0}^{2}\frac{r^{2}}{L^{2}}f(r){\mathcal{H}}^{\frac{2}{3}}-{\mathcal{H}}{\pi_{x}^{1}}^{2})-f(r){\pi_{y}^{1}}^{2}
\right]^{-\frac{1}{2}}.
\end{eqnarray}
As before, by using reality condition one can obtain,
\begin{equation}\label{s42}
{\pi_{y}^{1}}^{2}=\left[(\frac{f(r)}{{\mathcal{H}}}-\frac{r^{2}v^{2}}{L^{2}})
(T_{0}^{2}\frac{r^{2}}{L^{2}}{\mathcal{H}}^{\frac{2}{3}}-\frac{{\mathcal{H}}}{f(r)}{\pi_{x}^{1}}^{2})\right]_{r=r_{min}},
\end{equation}
where $r_{min}$ is turnaround point. One can check easily that
$r_{min}\geq r_{c}$ ($r_{c}$ is critical radius which introduced in
the section 3). If $\pi_{y}^{1}=0$, then $r_{min}= r_{c}$ and above
solutions are similar to single quark solution ($l=0$). Here, in
order the string have a chance of turning around smoothly, it
requires that $\frac{\partial y}{\partial
x}=\frac{y^{\prime}}{x^{\prime}}=\infty$ at $r_{min}$ [51]. So, it
necessary to have $\pi_{x}^{1}=0$. Therefore one can find
$\pi_{y}^{1}=\frac{T_{0}}{L}r_{min}{\mathcal{H}}^{\frac{1}{3}}(r_{min})\sqrt{\frac{f(r_{min})}{{\mathcal{H}}(r_{min})}
-\frac{r_{min}^{2}v^{2}}{L^{2}}}$.\\
In the second case we add a rotational motion with angular velocity
$\dot{\theta}$. Therefore the string may be described by the $x(r,
t)=vt+x(r)\sin\theta$ and $y(r, t)=y(r)\cos\theta$. So Fig. 1 shows
the configuration of rotating string. As we can see, $\theta(t)$ is
an angle with $Y$ axis. These solutions satisfy boundary conditions
$x(\infty, t)=vt\pm\frac{l}{2}\sin\theta$ and $y(\infty,
t)=\pm\frac{l}{2}\cos\theta$, where for $\theta=0$ reduce to the
boundary condition without rotational motion. Also, from our
conjecture we have another condition as
$\frac{y^{\prime}}{x^{\prime}}=\cot\theta$, which reduces to
$\frac{y^{\prime}}{x^{\prime}}\rightarrow\infty$ at the
$\theta\rightarrow0$ limit, which is agree with the first case.\\
\begin{tabular*}{2cm}{cc}
\includegraphics[scale=0.5]{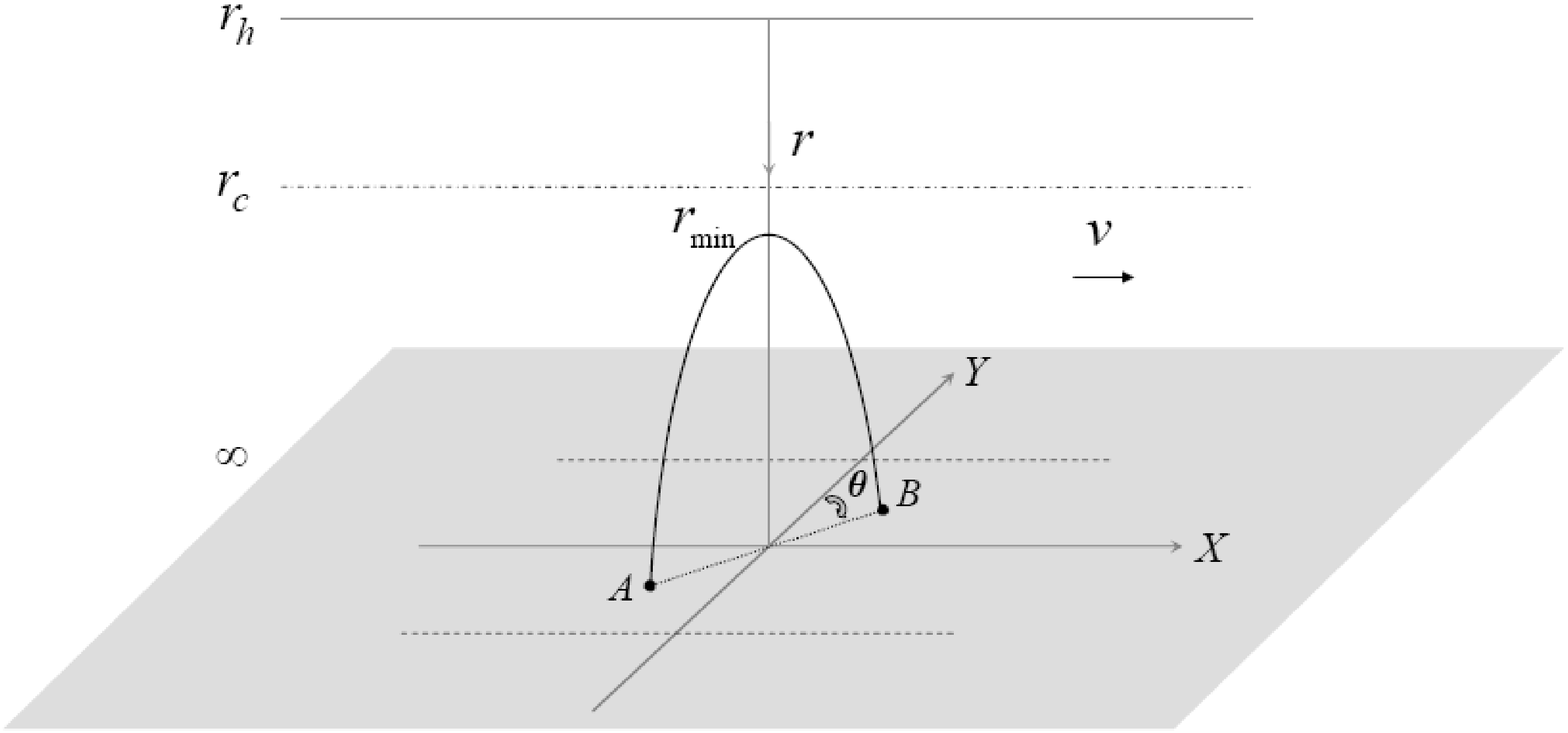}\\
\end{tabular*}\\
Figure 1: \emph{A rotating $\cap$ - shape string dual to a
$q\bar{q}$ pair that can be interpreted as a meson. $A$ and $B$
represent quark and antiquark with separating length $l$. The radial
coordinate $r$ varies from $r_h$ (black hole horizon radius) to
$r=r_{0}$ on $D$-brane. $r_c$ is a critical radius, obtained for
single quark solution, which the string can't penetrate beyond it
and $r_{min}\geq r_c$. $r_{min}=r_c$ is satisfied if points $A$ and
$B$ located at origin ($l=0$), in that case there is straight string
which is dual picture of single static quark. $\theta$ is assumed to
be the angle with $Y$ axis and the string center of mass moves along
$X$ axis with velocity
$v$.}\\\\\\\\\\
These boundary conditions can also satisfy with two separated string
which move at velocity $v$ along $X$ axis and simultaneously swing a
circle with radius $\frac{l}{2}$. Specifying these boundary
conditions doesn't lead to a unique solution for equation of motion,
so we should specify additional conditions for this motion. Here we
assume that the string is initially upright, move at velocity $v$
and rotates around its
center of mass.\\
Now by using above solutions in the equation (38) and solving
resulting equations with respect to $x^{\prime}$ and $y^{\prime}$
one can obtain following equations,
\begin{eqnarray}\label{s43}
A{x^{\prime}}^{2}+B{y^{\prime}}^{2}+Cx^{\prime}y^{\prime}+D&=&
0,\nonumber\\
A^{\prime}{x^{\prime}}^{2}+B^{\prime}{y^{\prime}}^{2}+C^{\prime}x^{\prime}y^{\prime}+D^{\prime}&=&
0,
\end{eqnarray}
where,
\begin{eqnarray}\label{s44}
A&=&R^{2}\sin^{2}\theta
\left[{\pi_{x}^{1}}^{2}\left(\frac{f(r)}{{\mathcal{H}}^{\frac{1}{3}}}
-{\mathcal{H}}^{\frac{2}{3}}y^{2}{\dot{\theta}}^{2}R^{2}\sin^{2}\theta\right)
-T_{0}^{2}{\mathcal{H}}^{\frac{2}{3}}R^{2}
\left({\mathcal{H}}^{\frac{1}{3}}y^{2}{\dot{\theta}}^{2}R^{2}\sin^{2}\theta
-\frac{f(r)}{{\mathcal{H}}^{\frac{2}{3}}}\right)^{2}\right],\nonumber\\
B&=&R^{2}\cos^{2}\theta\left[{\pi_{x}^{1}}^{2}\left(\frac{f(r)}{{\mathcal{H}}^{\frac{1}{3}}}
-{\mathcal{H}}^{\frac{2}{3}}(v+x\dot{\theta}\cos\theta)^{2}R^{2}\right)
-T_{0}^{2}{\mathcal{H}}^{\frac{4}{3}}y^{2}{\dot{\theta}}^{2}
R^{6}\sin^{2}\theta(v+x\dot{\theta}\cos\theta)^{2}\right],\nonumber\\
C&=&-2y\dot{\theta}R^{4}\sin^{2}\theta\left[{\pi_{x}^{1}}^{2}{\mathcal{H}}^{\frac{2}{3}}\cos\theta
+T_{0}^{2}{\mathcal{H}}
R^{2}\left({\mathcal{H}}^{\frac{1}{3}}y^{2}{\dot{\theta}}^{2}R^{2}\sin^{2}\theta
-\frac{f(r)}{{\mathcal{H}}^{\frac{2}{3}}}\right)\right](v+x\dot{\theta}\cos\theta),\nonumber\\
D&=&R^{2}{\pi_{x}^{1}}^{2}\frac{{\mathcal{H}}^{\frac{2}{3}}}{f(r)}
\left[\frac{f(r)}{R^{2}{\mathcal{H}}}
-y^{2}{\dot{\theta}}^{2}\sin^{2}\theta-(v+x\dot{\theta}\cos\theta)^{2}\right],\nonumber\\
A^{\prime}&=&R^{2}\sin^{2}\theta
\left[{\pi_{y}^{1}}^{2}\left(\frac{f(r)}{{\mathcal{H}}^{\frac{1}{3}}}
-{\mathcal{H}}^{\frac{2}{3}}y^{2}{\dot{\theta}}^{2}R^{2}\sin^{2}\theta\right)
-T_{0}^{2}{\mathcal{H}}^{\frac{4}{3}}y^{2}{\dot{\theta}}^{2}R^{6}
\sin^{2}\theta(v+x\dot{\theta}\cos\theta)^{2}\right],\nonumber\\
B^{\prime}&=&R^{2}\cos^{2}\theta\left[{\pi_{y}^{1}}^{2}\left(\frac{f(r)}{{\mathcal{H}}^{\frac{1}{3}}}
-{\mathcal{H}}^{\frac{2}{3}}(v+x\dot{\theta}\cos\theta)^{2}R^{2}\right)
-T_{0}^{2}{\mathcal{H}}^{\frac{2}{3}}R^{2}\left(R^{2}{\mathcal{H}}^{\frac{1}{3}}(v+x\dot{\theta}\cos\theta)^{2}
-\frac{f}{{\mathcal{H}}^{\frac{2}{3}}}\right)^{2}\right],\nonumber\\
C^{\prime}&=&-2y\dot{\theta}R^{4}\sin^{2}\theta\cos\theta\left[{\pi_{y}^{1}}^{2}{\mathcal{H}}^{\frac{2}{3}}
+T_{0}^{2}{\mathcal{H}}
R^{2}\left(R^{2}{\mathcal{H}}^{\frac{1}{3}}(v+x\dot{\theta}\cos\theta)^{2}
-\frac{f(r)}{{\mathcal{H}}^{\frac{2}{3}}}\right)\right](v+x\dot{\theta}\cos\theta),\nonumber\\
D^{\prime}&=&R^{2}{\pi_{y}^{1}}^{2}\frac{{\mathcal{H}}^{\frac{2}{3}}}{f(r)}
\left[\frac{f(r)}{R^{2}{\mathcal{H}}}
-y^{2}{\dot{\theta}}^{2}\sin^{2}\theta-(v+x\dot{\theta}\cos\theta)^{2}\right],
\end{eqnarray}
where we set $\frac{r}{L}=R$. We must note that the variable $C$ in
equations (43) and (44) is different with integration constant in
equation (11). Therefore from equations (43) one can obtain,
\begin{eqnarray}\label{s45}
x^{\prime}(r)&=&2\left[\frac{D(B-\frac{{\pi_{y}^{1}}^{2}}{{\pi_{x}^{1}}^{2}}B^{\prime})}{C^{2}-{C^{\prime}}^{2}
-4(BA-B^{\prime}A^{\prime})}\right]^{\frac{1}{2}},\nonumber\\
y^{\prime}(r)&=&2\left[\frac{D(A-\frac{{\pi_{y}^{1}}^{2}}{{\pi_{x}^{1}}^{2}}A^{\prime})}{C^{2}-{C^{\prime}}^{2}
-4(BA-B^{\prime}A^{\prime})}\right]^{\frac{1}{2}}.
\end{eqnarray}
Here, if the rotational motion vanishes ($\dot{\theta}=0$), from
equations (43) one can see that coefficients of
$x^{\prime}y^{\prime}$ vanish ($C=C^{\prime}=0$) and our solutions
recover the motion of quark-antiquark pair without rotation. In
order to obtain drag force we use reality condition and find a
relation between variable (44) as
$\frac{A}{A^{\prime}}=\frac{B}{B^{\prime}}=\frac{C}{C^{\prime}}=\frac{D}{D^{\prime}}=\frac{{\pi_{y}^{1}}^{2}}{{\pi_{x}^{1}}^{2}}$.
Then one can find two equations as, $C^{2}-4AB=0$ and
${C^{\prime}}^{2}-4A^{\prime}B^{\prime}=0$. These equations specify
$\pi_{x}^{1}$ and $\pi_{y}^{1}$ respectively. After some
calculations and simplifications we obtain,
\begin{eqnarray}\label{s46}
{\pi_{x}^{1}}^{2}&=&\frac{1}{2a}
\left[\pm\sqrt{b^{2}-4ac}-b\right],\nonumber\\
{\pi_{y}^{1}}^{2}&=&\frac{1}{2a^{\prime}}
\left[\pm\sqrt{{b^{\prime}}^{2}-4a^{\prime}c^{\prime}}-b^{\prime}\right],
\end{eqnarray}
where
\begin{eqnarray}\label{s47}
a&=&\prod_{i}(1+\frac{q_{i}}{r_{min}^{2}})^{\frac{1}{3}}\cos^{2}\theta(R_{min}^{2}\zeta+\xi\chi),\nonumber\\
b&=&T_{0}^{2}\prod_{i}(1+\frac{q_{i}}{r_{min}^{2}})^{\frac{2}{3}}\chi\left(2R_{min}^{4}\zeta+\cos^{2}\theta(R_{min}^{2}\xi\chi-\zeta)\right),\nonumber\\
c&=&T_{0}^{4}\prod_{i}(1+\frac{q_{i}}{r_{min}^{2}})R_{min}^{6}\sin^{2}\theta\zeta\chi^{2},\nonumber\\
a^{\prime}&=&\prod_{i}(1+\frac{q_{i}}{r_{min}^{2}})^{\frac{2}{3}}(R_{min}^{2}\zeta+\xi\chi),\nonumber\\
b^{\prime}&=&T_{0}^{2}R_{min}^{2}\xi\left(R_{min}^{4}\zeta\prod_{i}(1+\frac{q_{i}}{r_{min}^{2}})^{\frac{2}{3}}-2R_{min}^{2}\zeta-\prod_{i}(1+\frac{q_{i}}{r_{min}^{2}})^{\frac{2}{3}}\xi\chi\right),\nonumber\\
c^{\prime}&=&T_{0}^{2}\prod_{i}(1+\frac{q_{i}}{r_{min}^{2}})^{\frac{2}{3}}R_{min}^{6}\zeta\xi^{2}\left(\prod_{i}(1+\frac{q_{i}}{r_{min}^{2}})^{\frac{4}{3}}-T_{0}^{2}\right),
\end{eqnarray}
with,
\begin{eqnarray}\label{s48}
\zeta&=&\prod_{i}(1+\frac{q_{i}}{r_{min}^{2}})y^{2}{\dot{\theta}}^{2}R_{min}^{2}\sin^{2}\theta(v+x\dot{\theta}\cos\theta)^{2},\nonumber\\
\xi&=&\frac{f(r_{min})}{\prod_{i}(1+\frac{q_{i}}{r_{min}^{2}})^{\frac{1}{3}}}
-\prod_{i}(1+\frac{q_{i}}{r_{min}^{2}})^{\frac{2}{3}}(v+x\dot{\theta}\cos\theta)^{2}R_{min}^{2},\nonumber\\
\chi&=&\prod_{i}(1+\frac{q_{i}}{r_{min}^{2}})^{\frac{1}{3}}y^{2}{\dot{\theta}}^{2}R_{min}^{2}\sin^{2}\theta
-\frac{f(r_{min})}{\prod_{i}(1+\frac{q_{i}}{r_{min}^{2}})^{\frac{2}{3}}},
\end{eqnarray}
where $R_{min}=\frac{r_{min}}{L}$ and $r_{min}$ is the turnaround
point. The direct consequence of rotational motion is that drag
force is no longer constant. From equation (46) one can see that the
momentum densities
of string vary with respect to $x(r)$ and $y(r)$.\\
But this is not appropriate description of a meson. Indeed,
according to previous works [26, 27] the $q\bar{q}$ pair should be
close enough together and not moving too quickly. The presence of
functions $x(r)$ and $y(r)$ in relations (48) is consequence of
relativistic motion, which is not acceptable. On the other hand,
because of non-vanishing drag forces, it is expected that the
velocity of a $q\bar{q}$ pair decreases. So, we consider a moving
heavy $q\bar{q}$ pair with non-relativistic speed, which rotates by
angel $\theta=\omega t$ around the center of mass. Indeed this
situation is corresponding to motion of the heavy meson with large
spin. Actually, in the very large angular momentum limit, a
classical approximation is reliable. In this case, the angular
velocity of the string is very small. Therefore we are going to the
case of non-relativistic motion ($\dot{\theta}^{2}\rightarrow0$ and
$\dot{\theta}v\rightarrow0$). In that case $\zeta=c=c^{\prime}=0$
and we have,
\begin{eqnarray}\label{s49}
{\pi_{x}^{1}}^{2}&=&\frac{r_{min}^{2}}{L^{2}}T_{0}^{2} f(r_{min})
\mathcal{H}^{-\frac{1}{3}}(r_{min}),\nonumber\\
{\pi_{y}^{1}}^{2}&=&\frac{r_{min}^{2}}{L^{2}}T_{0}^{2}\left(f(r_{min})-\mathcal{H}(r_{min})\frac{r_{min}^{2}v^{2}}{L^{2}}\right)
\mathcal{H}^{-\frac{1}{3}}(r_{min}).
\end{eqnarray}
Now we assume that $v^{2}\rightarrow0$ and angular velocity is
infinitesimal constant ($\dot{\theta}=\omega\ll1$), and the
quark-antiquark pair rotate around origin. In that case we neglect
$\omega^{4}$ terms and obtain values of momentum densities as,
\begin{equation}\label{s50}
\pi_{x}^{1}=\pi_{y}^{1}=T_{0}\frac{r_{c}}{L}\frac{\left[1-\frac{\eta}{r_{c}^{2}}
+\frac{r_{c}^{2}}{L^{2}}{\prod_{i}(1+\frac{q_{i}}{r_{c}^{2}})}\right]^{\frac{1}{2}}}
{\prod_{i}(1+\frac{q_{i}}{r_{c}^{2}})^{\frac{1}{6}}}.
\end{equation}
In order to obtain non-zero components of momentum densities (49)
and (50) we should use negative sign in relations (46), therefore
correct sign in equations (46) is minus sign.\\
As we saw for the non-relativistic motion, we have constant drag
forces.
\section{Higher derivative corrections}
In this section, we want to calculate the effect of higher
derivative terms in the drag force on the single quark moving
through $\mathcal{N}$=2 supergravity thermal plasma. If we consider
the lowest order of string length then we can expand the effective
action in power of $T_{0}^{-1}$. So, the high order terms of
$T_{0}^{-1}$ is corresponding to the higher derivative terms. As we
know, the higher order corrections are depend to the black hole
physics [1, 41, 42, 52], and allow us to have better understanding
of $AdS$/CFT correspondence. It is known that [1, 41], the full
lagrangian including higher derivative terms is,
$\mathcal{L}=\mathcal{L}_{0}+\mathcal{L}_{R^{2}}+\cdots$. Here just
we consider the first order correction to the R-charged black hole
solution. In presence of the first order correction the solution (2)
should be modified as the following [42],
\begin{eqnarray}\label{s51}
f&=&f_{0}+c_{1}f_{1},\nonumber\\
H_{i}&=&h_{i0}+c_{1}h_{i1} \hspace{10mm} i=1, 2, 3,
\end{eqnarray}
where $f_{0}=1-\frac{\eta}{r^{2}}+\frac{r^{2}}{L^{2}}{\mathcal{H}}$
and  $h_{i0}=1+\frac{q_{i}}{r^{2}}$ and $c_{1}$ is constant small
parameter. The correction terms are obtained by [42],
\begin{eqnarray}\label{s52}
f_{1}&=&\frac{\eta^{2}}{96r^{6}h_{i0}}-\frac{5q_{i}(q_{i}+\eta)}{72L^{2}r^{4}},\nonumber\\
h_{i1}&=&-\frac{q_{i}(q_{i}+\eta)}{72r^{6}h_{i0}},
\end{eqnarray}
and one can obtain,
\begin{equation}\label{s53}
{\mathcal{H}}={\mathcal{H}_{0}}+c_{1}{\mathcal{H}_{1}}=\prod_{i=1}^{3}\left(1+\frac{q_{i}}{r^{2}}-\frac{c_{1}q_{i}(q_{i}+\eta)}{72r^{2}(r^{2}+q_{i})^{2}}\right).
\end{equation}
By using above corrections to the case of moving single quark with
constant speed $v$, one can obtain lagrangian density as the
following,
\begin{equation}\label{s54}
{\mathcal{L}}=-\frac{1}{({{\mathcal{H}_{0}}+c_{1}{\mathcal{H}_{1}}})^{\frac{1}{6}}}
\left[1-\frac{({\mathcal{H}_{0}}+c_{1}{\mathcal{H}_{1}})r^{2}}{(f_{0}+c_{1}f_{1})L^{2}}v^{2}
+\frac{(f_{0}+c_{1}f_{1})r^{2}}{L^{2}}{x^{\prime}}^{2}\right]^{\frac{1}{2}}.
\end{equation}
So, after some calculations similar to the section 3 one can find
momentum density as the following relation,
\begin{equation}\label{s55}
\pi_{x}^{1}=\left[(h_{10}+c_{1}h_{11})(h_{20}+c_{1}h_{21})(h_{30}+c_{1}h_{31})\right]^{\frac{1}{3}}vr_{c}^{2},
\end{equation}
where $r_{c}$ is the critical point which is root of the following
equation,
\begin{equation}\label{s56}
f_{0}+c_{1}f_{1}-r^{2}v^{2}({{\mathcal{H}_{0}}+c_{1}{\mathcal{H}_{1}}})=0.
\end{equation}
So, we can generalize ${x^{\prime}}^{2}$ as a following,
\begin{equation}\label{s57}
{x^{\prime}}^{2}=v^{2}L^{2}r_{h}^{4}\left[{{\mathcal{H}_{0}}+c_{1}{\mathcal{H}_{1}}}\right]_{r=r_{h}}^{\frac{1}{3}}
\frac{({{\mathcal{H}_{0}}+c_{1}{\mathcal{H}_{1}}})^{\frac{2}{3}}}{(f_{0}+c_{1}f_{1})r^{4}},
\end{equation}
and the corrected horizon radius in presence of higher derivative
terms is given by,
\begin{equation}\label{s58}
r_{h}=r_{0h}
\left[1+\frac{c_{1}}{24}\frac{\frac{{\mathcal{H}_{0}}^{\frac{4}{3}}}{L^{4}}(\sum
q_{i}^{2}-\frac{26r_{0h}^{2}}{3}\sum
q_{i}+3r_{0h}^{4})-\frac{2{\mathcal{H}_{0}}^{\frac{2}{3}}}{L^{2}}(\frac{13}{3}\sum
q_{i}-3r_{0h}^{2})+3}{24{\mathcal{H}_{0}}^{\frac{1}{3}}r_{0h}
\left[\frac{{\mathcal{H}_{0}}^{\frac{2}{3}}}{L^{2}}(\frac{1}{3}\sum
q_{i}-2r_{0h}^{2})-1\right]}\right],
\end{equation}
where $r_{0h}$ is the horizon radius without higher derivative
corrections which is obtained by the following equation,
$1-\frac{\eta}{r^{2}}+\frac{r^{2}}{L^{2}}\prod_{i=1}^{3}(1+\frac{q_{i}}{r^{2}})=0$.
If we assume that $q_{2}=q_{3}=0$ and $q=\eta\sinh^{2}\beta$ above equations reduce to [1].\\
Finally the energy and momentum densities along the string is
modified (due to higher derivative terms) by the following
equations,
\begin{equation}\label{s59}
\pi_{t}^{0}=-T_{0}\left[\frac{r_{c}^{4}}{r_{h}^{4}}
\frac{({{\mathcal{H}_{0}}+c_{1}{\mathcal{H}_{1}}})_{r=r_{c}}^{\frac{2}{3}}}
{({{\mathcal{H}_{0}}+c_{1}{\mathcal{H}_{1}}})_{r=r_{h}}^{\frac{1}{3}}}
\right]^{\frac{1}{2}}\left(1+v^{2}r_{h}^{2}({{\mathcal{H}_{0}}+c_{1}{\mathcal{H}_{1}}})_{r=r_{h}}^{\frac{1}{3}}\right)
\frac{({{\mathcal{H}_{0}}+c_{1}{\mathcal{H}_{1}}})^{\frac{1}{3}}}{(f_{0}+c_{1}f_{1})r^{2}},
\end{equation}
and
\begin{equation}\label{s60}
\pi_{x}^{0}=T_{0}\frac{vr_{c}^{4}}{L^{2}r_{h}^{4}}
\frac{({{\mathcal{H}_{0}}+c_{1}{\mathcal{H}_{1}}})_{r=r_{c}}^{\frac{2}{3}}}
{({{\mathcal{H}_{0}}+c_{1}{\mathcal{H}_{1}}})_{r=r_{h}}^{\frac{1}{3}}}
\frac{({{\mathcal{H}_{0}}+c_{1}{\mathcal{H}_{1}}})^{\frac{2}{3}}r^{2}}{(f_{0}+c_{1}f_{1})}.
\end{equation}
From the modified solution (52) one can see that, at the
near-extremal limit of the first order corrections have not any
effect to the lagrangian density and drag force, because at the
$\eta\rightarrow0$ limit, $f_{1}=h_{i1}=0$ and $\mathcal{H}=1$.
Actually if one consider higher order terms (second, third,...) at
the near-extremal limit there are no effect on the drag force,
because all of the correction terms are depend to the
non-extremality parameter [41].
\section{Conclusion}
In this paper we developed our previous papers about drag force of
moving quark and quark-antiquark pair through ${\mathcal{N}}$=2
supergravity thermal [1, 2]. We generalized our works to the case of
three non-zero charged black hole. Indeed, we considered a moving
quark and also a rotating quark-antiquark pair through
${\mathcal{N}}$=2 supergravity thermal plasma with three-charge
non-extremal black hole background, where $q_{1}\neq q_{2}\neq
q_{3}\neq0$. Furthermore, this paper has more discussions and
calculations than Ref. [2], such as effect of constant NSNS B-field,
effect of higher derivative corrections, and quark-antiquark
solution. Also we found some new results which
summarize here.\\
In the section 3 we concluded that the electromagnetic field
strength on the D-brane is necessary to keep the motion of string at
the constant velocity. Then in the section 5, we found that type of
this constant NSNS B-field should be electric field. Also in the
section 3, we discussed about special case of $q_{1}=q_{2}=q$ and
$q_{3}=0$, and found that for small velocity (non-relativistic case
with $v^{2}\ll1$) and near-extremal limit ($\eta\rightarrow1$) there
is the maximum value of drag force if $q=\frac{1}{8}$. In the case
of relativistic motion at the near-extremal limit we obtained
maximum value of drag force
coefficient.\\
We used the method of Refs. [1, 10] in the case of STU model
background with three non-zero charges and with different values For
the quasi-normal modes of curved string. We have shown that all of
our results are agree with Refs. [10, 11, 15, 53] at the
near-extremal limit. In fact, we calculated the drag force in the
${\mathcal{N}}$=2 supergravity which has non-extremal black hole.
But in the ${\mathcal{N}}$=4 SYM theory there are near-extremal
black hole. So it is expected that the near-extremal limit of the
${\mathcal{N}}$=2 supergravity theory would be
corresponding to the  ${\mathcal{N}}$=4 SYM theory.\\
The section of the quark-antiquark solutions has two pieces. The
first, was the simple extension of previous works such as [51] for
the case of ${\mathcal{N}}$=2 supergravity thermal plasma with three
non-zero black hole charges. The second, was consideration of
rotating quark-antiquark pair. For the first time, we calculated the
drag force of rotating quark-antiquark pair in the ${\mathcal{N}}$=4
SYM theory [29]. Now in STU model we have done the same calculations
and found the drag force for rotating $q\bar{q}$ pair with
non-relativistic motion have two constant components ($\pi_{x}$ and
$\pi_{y}$), while in the case of just linear motion it is found that
$\pi_{x}$ should be vanishes to have real motion. In summary it
should be noted that the ${\mathcal{N}}$=2 supergravity is an
unusual interpretation, and the usual one is ${\mathcal{N}}$=4 SYM
at $\mu\rightarrow0$ limit, so we have shown that there is not
difference at all
between the two calculations.\\
Finally in the last section by using results of the Ref. [42] we
considered effect of higher order corrections (first order only). We
represented modified horizon radius and generalized drag force of
single quark due to higher derivative terms.\\
In here there are many interesting problems, which we introduce some
of them as follows. There are other parameters more than the drag
force such as shear viscosity [54, 55, 56] and jet quenching
parameter [17, 18, 19, 20, 21, 22, 23, 24, 57] in the STU background
with non-zero charges. Other interesting problem is consideration of
more quark states such as four quarks in the baryon [58]. Also it is
interesting to check the possibility of extension our work to the
case of four-charge or 8-charge black hole in STU
model [44, 59, 60].\\\\
\textbf{Acknowledgments}\\\\
The authors would like to thank F. Larsen for giving some useful
motivation about STU model and C.P. Herzog for reading manuscript
and giving good comments. Also we would like to thank organizers of
Spring School on Superstring Theory and Related Topics, ICTP, 2009
and IPM String School and Workshop, Tehran, 2009.


\begin{thebibliography}{11}
\bibitem{P1}
J. Sadeghi and B. Pourhassan, " Drag force of moving quark at the
${\mathcal{N}} =2$ supergravity",JHEP12(2008)026, arXiv:0809.2668
[hep-th].
\bibitem{P2}
J. Sadeghi, M. R. Setare, B. Pourhassan and S. Hashmatian, "Drag
force of moving quark in STU background"Eur. Phys. J. \textbf{C
61}(2009) 527, arXiv:0901.0217 [hep-th].
\bibitem{P3}
J. M. Maldacena, "The large N limit of superconformal field theories
and supergravity", Adv. Theor. Math. Phys. \textbf{2} (1998) 231.
\bibitem{P4}
E. Witten, "Anti-de Sitter space and holography", Adv. Theor. Math.
Phys. \textbf{2} (1998) 253.
\bibitem{P5}
J. H. Schwart, "Introduction to M Theory and $AdS$/CFT Duality",
Lecture Notes in Physics, Volume 525(1999), hep-th/9812037.
\bibitem{P6}
M. R. Douglas and S. Randjbar-Daemi, "Two Lectures on $AdS$/CFT
correspondence" hep-th/9902022.
\bibitem{P7}
J. L. Petersen, "Introduction to the Maldacena Conjecture on
$AdS$/CFT", Int.J.Mod.Phys. \textbf{A14} (1999) 3597.
\bibitem{P8}
Horatiu Nastase, "Introduction to AdS-CFT", arXiv:0712.0689v2
[hep-th].
\bibitem{P9}
Igor R. Klebanov, "TASI Lectures: Introduction to the AdS/CFT
Correspondence", arXiv:hep-th/0009139.
\bibitem{P10}
C. P. Herzog, A. Karch, P. Kovtun, C. Kozcaz, and L. G. Yaffe,
"Energy loss of a heavy quark moving through ${\mathcal{N}} =4$
supersymmetric Yang-Mills plasma" JHEP \textbf{0607} (2006) 013,
arXiv: hep-th/0605158.
\bibitem{P11}
C. P. Herzog, "Energy loss of heavy quarks from asymptotically $AdS$
geometries", JHEP \textbf{0609} (2006) 032, arXiv: hep-th/0605191.
\bibitem{P12}
S. S. Gubser, "Drag force in $AdS$/CFT", Phys. Rev. \textbf{D74}
(2006) 126005.
\bibitem{P13}
E. Nakano, S. Teraguchi and W. Y. Wen, "Drag Force, Jet Quenching,
and $AdS$/QCD", Phys. Rev. \textbf{D75} (2007) 085016.
\bibitem{P14}
E. Caceres and A. Guijosa, "Drag force in charged ${\mathcal{N}} =4$
SYM plasma". JHEP \textbf{0611} (2006) 077.
\bibitem{P15}
T. Matsuo, D. Tomino and W. Y. Wen, "Drag force in  SYM plasma with
$B$ field from $AdS$/CFT", JHEP \textbf{0610} (2006) 055.
\bibitem{P16}
J. F. Vazquez-Poritz, "Drag force at finite 't Hooft coupling from
$AdS$/CFT", arXiv: hep-th/0803.2890.
\bibitem{P17}
H. Liu, K. Rajagopal and U. A. Wiedemann, "Calculating the jet
quenching parameter from $AdS$/CFT", Phys. Rev. Lett. \textbf{97}
(2006) 182301.
\bibitem{P18}
J. F. Vazquez-Poritz, " Enhancing the jet quenching parameter from
marginal deformations", arXiv: hep-th/0605296.
\bibitem{P19}
E. Caceres and A. Guijosa, "On drag forces and jet quenching in
strongly coupled plasmas", JHEP \textbf{0612} (2006) 068.
\bibitem{P20}
F. L. Lin and T. Matsuo, "Jet quenching parameter in medium with
chemical potential from $AdS$/CFT", Phys. Lett. \textbf{B641} (2006)
45.
\bibitem{P21}
S. D. Avramis and K. Sfetsos, "Supergravity and the jet quenching
parameter in the presence of $R$-charge densities", JHEP
\textbf{0701} (2007) 065.
\bibitem{P22}
N. Armesto, J. D. Edelstein and J. Mas, "Jet quenching at finite 't
Hooft coupling and chemical potential from $AdS$/CFT", JHEP
\textbf{0609} (2006) 039.
\bibitem{P23}
J. D. Edelstein and C. A. Salgado, "Jet quenching in heavy Ion
collisions from $AdS$/CFT", AIPConf.Proc.1031(2008)207, arXiv:
hep-th/0805.4515.
\bibitem{P24}
K. B. Fadafan, "Medium effect and finite 't Hooft coupling
correction on drag force and Jet Quenching Parameter", arXiv:
hep-th/0809.1336.
\bibitem{P25}
K. Peeters, J. Sonnenschein, M. Zamaklar, "Holographic melting and
related properties of mesons in a quark gluon plasma", Phys.Rev.
\textbf{D74} (2006) 106008.
\bibitem{P26}
H. Liu, K. Rajagopal, U. A. Wiedemann, "An AdS/CFT calculation of
screening in a hot wind", Phys.Rev.Lett.\textbf{98}(2007)182301,
arXiv: hep-ph/0607062.
\bibitem{P27}
M. Chernicoff, J. A. Garcia and A. Guijosa, "The Energy of a Moving
Quark-Antiquark Pair in an N=4 SYM Plasma", JHEP \textbf{0609}
(2006) 068.
\bibitem{P28}
J. Erdmenger, N. Evans, I. Kirsch, and E.J. Threlfall, "Mesons in
gauge/gravity duals", Eur. Phys. J. A \textbf{35} (2008) 81.
\bibitem{P29}
J. Sadeghi, B. Pourhassan, S. Heshmatian, "Rotating heavy meson in a
N=4 SYM plasma and AdS/CFT", arXiv:0812.4816 [hep-th].
\bibitem{P30}
M. Chernicoff, J. A. Garcia and A. Guijosa, "The Energy of a Moving
Quark-Antiquark Pair in an N=4 SYM Plasma", JHEP \textbf{0609}
(2006) 068.
\bibitem{P31}
R. Casero, A. Paredes, J. Sonnenschein, "Fundamental matter, meson
spectroscopy and non-critical string/gauge duality",
JHEP\textbf{0601}(2006)127, arXiv: hep-th/0510110.
\bibitem{P32}
M. Kruczenski, D. Mateos, R. C. Myers and D. J. Winters, "Meson
spectroscopy in AdS/CFT with flavour", JHEP \textbf{0703}, (2003)
049, arXiv: hep-th/0304032.
\bibitem{P33}
M. Kruczenski, L. A. Pando Zayas, J. Sonnenschein and D. Vaman,
"Regge trajectories for mesons in the holographic dual of large-N(c)
QCD, JHEP \textbf{0506}, (2005) 046, arXiv: hep-th/0410035.
\bibitem{P34}
K. Bitaghsir Fadafan, H. Liu, K. Rajagopal and U. Achim
Wiedemann,"Stirring Strongly Coupled Plasma",
Eur.Phys.J.\textbf{C61}(2009)553, arXiv: hep-th/0809.2869.
\bibitem{P35}
K. B. Fadafan, "$R^{2}$ curvature-squared correction on drag force",
JHEP12(2008)051, hep-th/0803.2777
\bibitem{P36}
Davide Gaiotto, Juan Maldacena, "The gravity duals of
$\mathcal{N}$=2 superconformal field theories", arXiv:0904.4466
[hep-th]
\bibitem{P37}
D. Gaiotto, "N=2 dualities", arXiv: 0904.2715 [hep-th].
\bibitem{P38}
K. Behrndt, A. H. Chamseddine and W. A. Sabra, "BPS black holes in
${\mathcal{N}} =2$ five dimensional $AdS$ supergravity", Phys. Lett.
\textbf{B442} (1998) 97.
\bibitem{P39}
K. Behrndt, M. Cvetic  and W. A. Sabra, "Non-extreme black holes of
five dimensional ${\mathcal{N}} =2$ $AdS$ supergravity", Nucl. Phys.
\textbf{B553} (1999) 317.
\bibitem{P40}
A. C. Cadavid, A. Ceresole, R. D'Auria, and S. Ferrara,
"Eleven-dimensional supergravity compactified on Calabi-Yau three
folds", Phys. Lett. \textbf{B357} (1995) 76, arXiv: hep-th/9506144.
\bibitem{P41}
J. T. Liu and P. Szepietowski, " Higher derivative corrections to
$R$-charged $AdS_{5}$ black hole and field redefinitions", arXiv:
hep-th/0806.1026.
\bibitem{P42}
Sera Cremonini, Kentaro Hanaki, James T. Liu, Phillip Szepietowski,
"Black holes in five-dimensional gauged supergravity with higher
derivatives", arXiv: hep-th/0812.3572.
\bibitem{P43}
Alex Buchel, "Shear viscosity of CFT plasma at finite coupling",
Phys.Lett.\textbf{B665}(2008)298, arXiv:0804.3161 [hep-th].
\bibitem{P44}
Vijay Balasubramanian, Finn Larsen, "On D-Branes and Black Holes in
Four Dimensions", Nucl.Phys. \textbf{B478} (1996) 199-208, arXiv:
hep-th/9604189.
\bibitem{P45}
D. T. Son, A. O. Starinets, "Hydrodynamics of $R$-charged black
holes", JHEP \textbf{0603} (2006) 052.
\bibitem{P46}
G. T. Horowitz and V. E. Hubeny, "Quasinormal modes of AdS black
holes and the approach to thermal equilibrium", Phys. Rev. \textbf{D
62} (2000) 024027.
\bibitem{P47}
T. R. Govindarajan and V. Suneeta, "Quasi-normal modes of AdS black
holes: A superpotential approach", Class. Quant. Grav. \textbf{18}
(2001) 265.
\bibitem{P48}
D. Birmingham, "Choptuik scaling and quasinormal modes in the
AdS/CFT correspondence", Phys. Rev. \textbf{D 64} (2001) 064024.
\bibitem{P49}
D. Birmingham, I. Sachs and S. N. Solodukhin, "Conformal field
theory interpretation of black hole quasi-normal modes", Phys. Rev.
Lett. \textbf{88} (2002) 151301.
\bibitem{P50}
A. O. Starinets, "Quasinormal modes of near extremal black branes",
Phys. Rev. \textbf{D 66} (2002) 124013.
\bibitem{P51}
M. Chernicoff, J. A. Garcia and A. Guijosa, "The Energy of a Moving
Quark-Antiquark Pair in an N=4 SYM Plasma", JHEP \textbf{0609}
(2006) 068, arXiv: hep-th/0607089.
\bibitem{P52}
M.F. Paulos, "Higher derivative terms including the Ramond-Ramond
five-form", JHEP\textbf{0810}(2008) 047, arXiv:0804.0763 [hep-th].
\bibitem{P53}
M. Chernicoff, J. A. Garcia and A. Guijosa, "Energy Loss of Gluons,
Baryons and k-Quarks in an $\mathcal{N}$=4 SYM Plasma",
JHEP\textbf{0702}(2007) 084, arXiv: hep-th/0611155.
\bibitem{P54}
A. Buchel and J. T. Liu, " Universality of the shear viscosity in
supergravity", Phys.Rev.Lett. \textbf{93} (2004) 090602.
\bibitem{P55}
K. Meada, M. Natsuume, and T. Okamura, "Viscosity of gauge theory
plasma with a chemical potential from $AdS$/CFT correspondence",
Phys.Rev. \textbf{D73} (2006) 066013.
\bibitem{P56}
J. Mas, "Shear viscosity from $R$-charged $AdS$ black holes", JHEP
\textbf{0603} (2006) 016.
\bibitem{P57}
Philip C. Argyres, Mohammad Edalati, Justin F. Vazquez-Poritz,
"Spacelike strings and jet quenching from a Wilson loop"
JHEP0704,(2007) 049, and  "Lightlike Wilson loops from AdS/CFT"
JHEP0803 (2008) 071.
\bibitem{P58}
C. Krishnan, "Baryon Dissociation in a Strongly Coupled Plasma",
JHEP0812(2008) 019, arXiv:0809.5143 [hep-th].
\bibitem{P59}
Finn Larsen, "Anti-DeSitter Spaces and Nonextreme Black Holes",
arXiv: hep-th/9806071.
\bibitem{P60}
Eric Gimon, Finn Larsen, Joan Simon, "Constituent Model of Extremal
non-BPS Black Holes", JHEP \textbf{0907}(2009)052, arXiv:0903.0719
[hep-th].
\end{thebibliography}
\end{document}